# Nanoscale changes with low temperature annealing inside composite optical fibres

Wen Liu,[1,2] John Canning,[1,*] Kevin Cook,[1] and Cicero Martelli[3]

[1]interdisciplinary Photonics Laboratories (iPL), School of Chemistry, The University of Sydney, NSW 2006, Australia
[2]College of Optoelectronic Science and Technology, National University of Defense Technology, Changsha, Hunan 410073, China
[3]Graduate School of Electrical Engineering and Applied Computer Science, Federal University of Technology-Paraná, Curitiba PR 80230-901, Brazil
[*]john.canning@sydney.edu.au

**Abstract:** Nanoscale glass milling with mild thermal annealing is proposed and shown to occur within a tri-material composite "W" profile optical fibre. Evidence of glass relaxation with annealing is inferred directly through measured diameter changes in the core and inner cladding of the long period gratings (LPGs) ($\Delta\phi_{core} = (0.13 \pm 0.05)$ μm; $\Delta\phi_{inner\ cladding} = -(0.5 \pm 0.2)$ μm) using scanning electron microscopy (SEM) before and after annealing at T ~300 °C for 1 hour. Large reductions in the magnitudes of both the temperature and strain coefficients are observed after annealing. The temperature sensitivity drops from $d\lambda/dT = -(111.2 \pm 2.4)$ pm/K to $d\lambda/dT = -8.75 \times 10^{-2}$ pm/K and the strain sensitivity decreases from $d\lambda/d\varepsilon = -(0.11 \pm 0.13)$ pm/με to $d\lambda/d\varepsilon = 2.21 \times 10^{-2}$ pm/με. The fabrication of LPGs using 193 nm radiation is shown to produce measurable increases in the core dimensions within tri-material composite fibres whereas no changes are observed under similar conditions for commercial bi-material single mode fibre.

INTRODUCTION

Nanoscale glass milling using lasers was recently proposed as a new approach to engineering photonic components, highlighted by the use of UV inscription with thermal annealing [1–3]. The archetype demonstration is fibre Bragg grating (FBG) "regeneration" [28]. By annealing at high temperatures (typically > 800 °C) in the presence of hydrogen (H), it is possible to regenerate thermally stabilised gratings that can withstand temperatures potentially up to 1450 °C [4]. The primary origin for this dramatic improvement in stability is mechanical relaxation where changes in the processing history imparted by UV are locked into the glass, hydrogen creating a background pressure enhanced by the periodic formation of OH which changes interfacial/internal pressures periodically. These lead to subsequent structural relaxation during thermal annealing. The mechanical nature of this process was demonstrated by He regeneration [5]. There is one additional aspect, therefore, in this work which has not been investigated to date but is implicit in such relaxations – subsequent volume changes in the glass and the dimensional changes that must accompany them. Ordinarily, such changes in glass are familiar to glass blowers and smithers where high temperature annealing of bulk glass leads to structural changes and changes in fictive, $T_f$, and transition temperatures, $T_g$. Lower temperature regeneration has been demonstrated in so-called type 1$n$ FBGs [6], the earliest form of regeneration without H. In this case laser induced thermal annealing regenerates the grating in the germanosilicate core [7], reflecting the importance of glass composition in both lowering $T_g$ and temperatures where thermally induced relaxation occurs.

In this paper, we show that there are significant changes even for mild annealing at temperatures of only 200 °C or 300 °C, outside of the regeneration process and far below silica transition and fictive temperatures. These are standardised annealing temperatures for stabilizing gratings over a finite period where no changes are normally expected – their observation implies a radical rethink of annealing of glass more generally. The test bed for this are long period gratings (LPGs) – by virtue of coupling between core and cladding modes LPGs offer ultra-sensitive but straightforward measurement of such changes not possible by any other means.

Conventional LPGs are extremely promising for photonic filtering [8] and sensing applications [9,10], but suffer badly from cross-sensitivity between applied temperature and strain, requiring significant and elaborate packaging. To achieve stable filtering, LPGs intrinsically insensitive to strain or temperature are needed in most applications. Bhatia *et al.* [11] demonstrated insensitive LPGs by selecting an appropriate periodicity. Temperature coefficients d$\lambda$/d$T$ ~ -1.8 pm/K were achieved in single mode fibre (SMF 28) using a pitch $\Lambda$ = 40 μm – these are an order lower than that for longer pitch LPGs ($\Lambda$ > 100 um) [9]. On the other hand, longer periods exhibit lower strain coefficients – an LPG with $\Lambda$ = 340 μm produced d$\lambda$/d$\varepsilon$ ~ 0.04 pm/με. Others have optimised dopants such as Ge and B to lower d$\lambda$/d$T$ [12,13], whilst others still have removed them altogether in single-material structured optical fibres, reducing d$\lambda$/d$T$ further still [14,15]. The method of producing LPGs can also influence their performance – a $CO_2$ laser used to write an LPG in a structured optical fibre reduced strain sensitivity to d$\lambda$/d$\varepsilon$ = -0.192 pm/με [16]. In the following paper, we identify and explore the physical relaxation mechanisms that are common across such data. In doing so we show how both appropriate fibre design and thermal annealing can enhance nanoscale changes in the core and inner cladding diameters to produce environmentally insensitive LPGs just with simple annealing. The model is supported by simulation and the role of design is shown by comparing tri-material composite fibres (those with an inner cladding, so-called "W" fibres reflecting their index profile) with bi-material composite glass SMF 28 fibre. Longitudinal changes with annealing can be ignored because they will be less significant given the grating length relative to the fibre width. Instead, sufficient volume changes with mild annealing can be achieved to alter the dispersion properties of the fibre accordingly and optimise coupling as a function of both temperature and strain. This LPG "optical test bench" opens up a new and broader approach to nanoscale engineering and fine tuning technology for all composites and waveguides as well as providing insight into the science of glass relaxation.

RESPONSE OF LPG PRIOR TO ANNEALING

The strain sensitivity of an LPG includes the contributions from material and waveguide properties. For LPGs with periods larger than $\Lambda$ > 100 μm, the material and waveguide contributions are usually negative and positive respectively [9,10]. With careful design, it is feasible to offset the two contributions to achieve strain insensitivity – the origins of this have yet to be identified and the process is likely not optimised.

*2.1 W Fibre Configuration*

A custom-designed dual-cladding, tri-material composite "W" profile single mode fibre (W fibre) was drawn from its preform fabricated using modified chemical vapor deposition (MCVD). The low loss, enhanced photosensitivity design is matched to bi-material SMF28 fibre. Fig. 1 and Table 1 show refractive index and composition of the preform respectively.

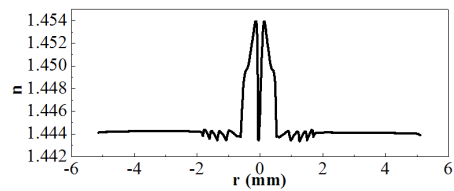

Fig. 1. Refractive index, *n*, profile of the W fibre preform.

**Table 1. Composition of the W fibre preform**

|  | $SiO_2$ (mol. %) | $B_2O_3$ (mol. %) | $GeO_2$ (mol. %) | $P_2O_5$ (mol. %) | F (mol. %) |
|---|---|---|---|---|---|
| Core | 47 | 20 | 33 | / | / |
| Inner cladding | >85 | / | / | 11 | <5 |
| Outer cladding | 100 | / | / | / | / |

In order to characterize the fibre cross-section with sufficient resolution, careful etching with HF ([HF] ~ 5%, etch rate ~ 0.1 μm/min, diluted in deionized water for 5 min) created topological contrast between core, inner cladding and outer cladding; such etching follows a 1:1 correspondence with chemical composition, where doped silica etches faster than undoped and is insensitive to existing or induced stresses [17]. A scanning electron microscope (SEM) was used to measure the cross-sectional dimensions of the drawn fibre and all measurements were repeated four times to ensure reproducibility and reliability. The average diameter of the core, the inner cladding and the outer cladding of the pristine W fibre were measured to be $\phi_{core}$ = (8.85 ± 0.05) μm, $\phi_{inner\ cladding}$ = (40.3 ± 0.2) μm, and $\phi_{outer\ cladding}$ = (124.8 ± 0.7) μm respectively. Compared with the unetched samples, the average outer diameter of the etched samples was 0.5 μm smaller, which was taken into account in the measurements; this variation is the main source of variability.

## 2.2 Seed LPG Fabrication

The output of a 193 nm ArF laser ($f_{pulse}$ ~ 50 mJ/cm$^2$; $f_{cum}$ ~ 225 J/cm$^2$; repetition rate $RR$ = 30 Hz; pulse duration $\tau_w$ = 15 ns) was used to inscribe LPGs through an amplitude mask (period $\Lambda$ ~ 325 μm, grating length $L$ ~ 10 mm). The transmission spectrum of the LPG was measured using a white light source and an optical spectrum analyser (OSA) over the wavelength range $\Delta\lambda$ = (1000 – 1700) nm. Two transmission rejection bands were observed at $\lambda_{LPG1}$ = 1550 nm and $\lambda_{LPG2}$ = 1400 nm, with rejection strengths $R_1$ ~ 15 dB and $R_2$ ~ 10 dB.

Four seed LPGs in the W fibre were cleaved at irradiated regions, etched identically to the pristine samples and measured under SEM. The average diameter of the core, the inner cladding and the outer cladding were $\phi_{core}$ = (8.97 ± 0.05) μm, $\phi_{inner\ cladding}$ = (40.4 ± 0.2) μm, and $\phi_{outer\ cladding}$ = (124.7 ± 0.7) μm. There is no change within experimental error from the pristine fibre dimensions except the small increase of the core diameter. Using these measured diameters and considering the errors, the transmission spectrum of one typical LPG was simulated by transfer matrix method (TMM) (Fig. 2) [18–20]. The simulation shows a slightly smaller separation between the two transmission rejection bands compared to the experimental results within the determined error range. This difference may be attributed to actual differences in fibre refractive index, $n$, including the stress-optic contribution introduced by drawing, to that assumed by simply calling the preform. Both notches disappeared when refractive index matching gel was placed on the grating indicating that the responsible modes are bound by the outer cladding. Two transmission rejection bands are observed and identified to correspond to coupling between the core mode and the cladding modes LP(0, 3) and LP(0, 5). Average index values are $n_{core}$ ~ 1.45077, $n_{inner\ cladding}$ ~ 1.44390 and $n_{outer\ cladding}$ ~ 1.44403.

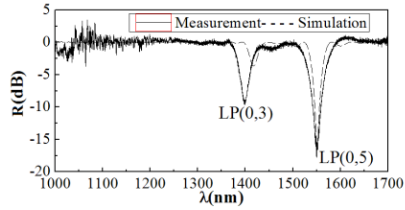

Fig. 2. Transmission spectrum of the W fibre LPG.

To establish the role of irradiation, four samples of W fibre ($L$ = 10 mm) were irradiated for $t$ = 2.5 min with bulk UV exposure (amplitude mask removed), using the same conditions for LPG fabrication. The dimensions were measured as $\phi_{core}$ = (8.88 ± 0.05) μm, $\phi_{inner\ cladding}$ = (40.7 ± 0.2) μm and $\phi_{outer\ cladding}$ = (124.6 ± 0.7) μm. These bulk exposures show a diameter increase of the inner cladding after irradiation whereas the core shows no increase within error.

## 2.3 Seed LPG Characterization

The temperature dependence was characterized by placing the LPGs in a computer controlled oven and monitoring transmission spectra over $T$ ~ 20 to 200 °C. The wavelength shift, $\Delta\lambda$, and the rejection strength, $R$, of the main transmission rejection band are plotted in Fig. 3. A near linear dependence, with a coefficient of $d\lambda/dT$ = -(111.2 ± 2.4) pm/K, is observed in Fig. 3 (a). The rejection band strength is stable at different temperatures as shown in Fig. 3 (b) where the measured transmitted power variation $\Delta R = \pm 0.2$ dB is the power fluctuation and drift of the white light source convolved with the resolution of the OSA.

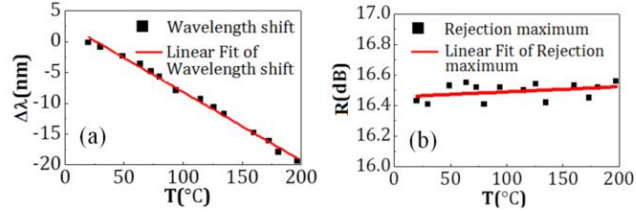

Fig. 3. Temperature sensitivity of the W fibre LPG: (a) wavelength shift, $\Delta\lambda$, and (b) rejection strength, $R$, versus temperature, $T$.

Strain sensitivity was measured using a translation stage by fixing one end of the LPG while stretching the other at an ambient temperature $T = 20$ °C (Fig. 4). Similar to the temperature dependence, a linear curve fit yields a linear but negative strain coefficient of $d\lambda/d\varepsilon = -(0.11 \pm 0.13)$ pm/με up to almost 1000 με. Throughout these tests, the W fibre LPG exhibited strain insensitivity with considerable temperature sensitivity, ideal for temperature measurements in the presence of random axial strain.

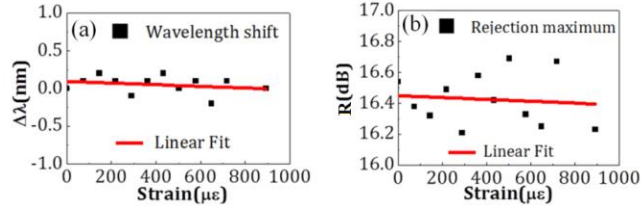

Fig. 4. Strain sensitivity of the W fibre LPG: (a) wavelength shift, $\Delta\lambda$, and (b) rejection strength, $R$, versus strain.

RESPONSE OF LPG WITH ANNEALING

The waveguide contribution to temperature dependence is positive [10] so the observation of a negative thermo-optic coefficient points to a material contribution, likely through existing stress. B-doped glasses are known to exhibit a negative thermo-optic coefficient of $dn/dT = -35.0 \times 10^{-6}$/K [12], but in this system when a normal FBG is inscribed, the thermo-optic coefficient is positive, demonstrating both waveguide and compositional changes are important. We predicted that annealing the fibre should affect both, as removing stresses should lead to glass relaxation which in turn should lead to greater waveguide dimensional changes. To this end, the LPGs were subjected to further annealing.

*3.1 Thermal Annealing Process*

The LPG was placed in the oven described earlier. The temperature was set to $T = 300$ °C and held there for 1 hour before cooling down. The transmission spectrum was measured continuously and the LPG wavelength tracked (Fig. 5(a)). The transmission rejection bands shifted smoothly with the two corresponding cladding modes, LP(0, 3) and LP(0, 5); no mode hopping was detected. The main rejection band remained stable and behaved as expected with increasing temperature, exhibiting a negative temperature coefficient. When the temperature was kept at 330 °C, the main rejection band reduced in strength (Fig. 5(b)), moving to shorter wavelengths during cooling.

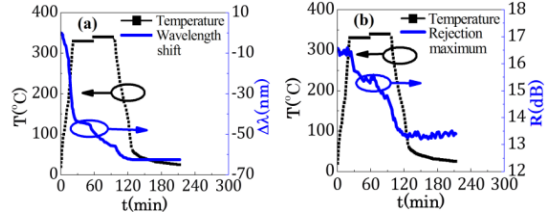

Fig. 5. The temperature profile and the evolution of the main rejection band of the W fibre LPG during the post-annealing process: (a) wavelength shift, $\Delta\lambda$, and (b) rejection strength, $R$.

Identical SEM analysis to that described earlier for the seed gratings was carried out, and four annealed LPGs were similarly etched and measured. The average diameters of the core, the inner cladding and the outer cladding were $\phi_{core} = (9.10 \pm 0.05)$ μm, $\phi_{inner\ cladding} = (39.9 \pm 0.2)$ μm, and $\phi_{outer\ cladding} = (124.6 \pm 0.7)$ μm respectively. Based on the results, the transmission spectrum of the annealed LPG was again simulated (Fig. 6). A substantially improved match is obtained consistent with mainly nanoscale waveguide dimension changes as the determinant of LPG spectra after annealing. The improved fitting supports a stress contribution in the index profile being responsible for the LPG prior to annealing.

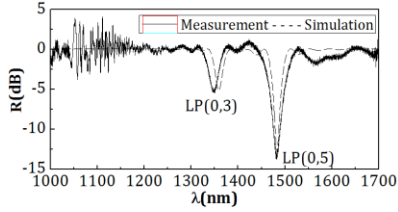

Fig. 6. Experimental and simulated results for the transmission spectrum of the annealed W fibre LPG.

*3.2 Annealed LPG characterization*

After post-annealing, $d\lambda/dT \sim 0$ when $T < 200$ °C. In order to confirm this temperature insensitivity a repeatability study was conducted with 4 LPGs, and each was subjected to 10 successive temperature ramps from $T = 20$ °C to $T = 200$ °C. The wavelength shift, $\Delta\lambda$, of the main rejection band for one LPG is shown in Fig. 7 (a) – an extremely small temperature coefficient of $d\lambda/dT = (1.3 \pm 0.3)$ pm/K is obtained. The strength of the main rejection band stabilised at $R \sim (13.4 \pm 0.2)$ dB (Fig. 7 (b)) throughout the cycles. The strain sensitivity and the corresponding strength are shown in Fig. 8. Compared with the LPGs before annealing, the value of the strain coefficient after annealing decreased slightly to $d\lambda/d\varepsilon = (0.02 \pm 0.11)$ pm/με, with strength variations of $\Delta R \sim \pm 0.2$ dB. The annealed LPGs are now simultaneously temperature and strain insensitive. For comparison, LPGs in a bi-material composite SMF 28 fibre (without comparable inner cladding) were fabricated using the same amplitude mask and annealed over 300 °C following the same procedure. There were four rejection bands in transmission over $\Delta\lambda = (1000 - 1700)$ nm – the main rejection band is located at $\lambda_{LPG}= 1550$nm, with $R \sim 7$ dB. Before and after annealing, $d\lambda/dT = (94.6 \pm 2.1)$ pm/K and $(100.9 \pm 1.1)$ pm/K, and $d\lambda/d\varepsilon = (0.91 \pm 0.12)$ pm/με and $(1.00 \pm 0.15)$ pm/με respectively (Fig. 9). In contrast with the W fibre, post-annealing did not induce significant differences for SMF 28 LPGs. Rather, the sensitivity increases slightly.

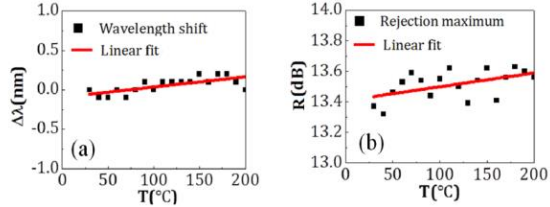

Fig. 7. Temperature sensitivity of the annealed W fibre LPG: (a) wavelength shift, Δλ, and (b) transmission rejection, R, versus temperature, T.

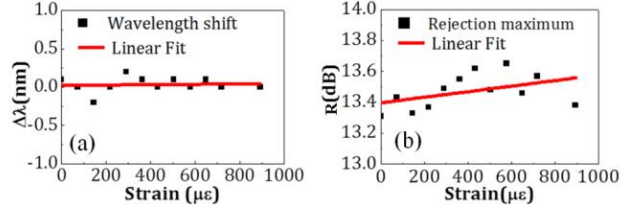

Fig. 8. Strain sensitivity of the annealed W fibre LPG: (a) wavelength shift, Δλ, and (b) transmission rejection, R, versus strain.

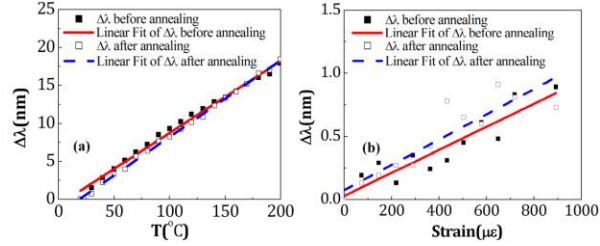

Fig. 9. Sensitivities of the SMF 28 LPG before and after annealing. (a) temperature sensitivity, and (b) strain sensitivity.

## MEASUREMENT OF NANOSCALE CHANGES

Fig. 10 summarises the W fibre LPG diameters before and after annealing. The dimensional changes within the W fibre LPG after annealing are clear: the core increases by $\Delta\phi_{core}$ ~ 0.13 μm whilst the inner cladding decreases by $\Delta\phi_{inner\ cladding}$ ~ -0.5 μm, pointing to a decrease in core index and an increase in inner cladding index. Compared to the softer inner region, the outer cladding remains stiff and within resolution no changes are detected. The experimental results and SEM images for the LPG gratings show the effect of post-annealing. An LPG has a long period so it would seem unlikely that there is a long range periodic effect involved. They are consistent with annealing through stress relaxation and subsequent structural and volume change that can and does, in the case of LPGs, impact waveguide properties. In order to confirm a change in waveguide local dimensions due to annealing, , four pristine samples and four samples of both the W fibre and SMF28 that have undergone bulk UV exposure were also annealed. All the samples were etched as before prior to SEM analysis. The measured quantities are shown in Table 2.

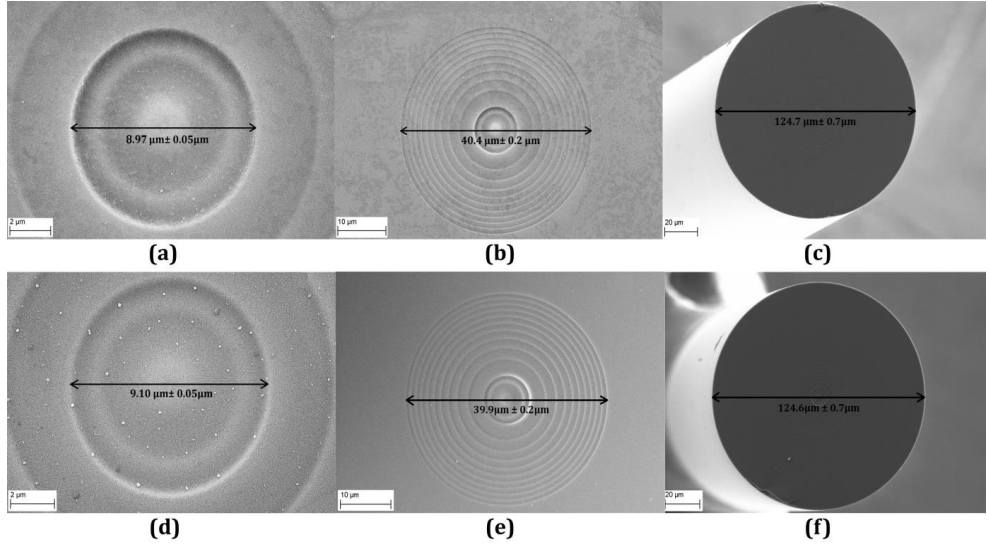

Fig. 10. The images of etched W fibre LPG samples before and after annealing under SEM: (a) core region before annealing; (b) inner cladding before annealing; (c) outer cladding before annealing; (d) core region after annealing; (e) inner cladding after annealing; (f) outer cladding after annealing.

**Table 2. Average diameters of different fibre sample**

|  | Core (μm) | Inner cladding (μm) | Outer cladding (μm) |
|---|---|---|---|
| Pristine Un-annealed W fibre | 8.85 ± 0.05 | 40.3 ± 0.2 | 124.8 ± 0.7 |
| Pristine Annealed W fibre | 8.87 ± 0.05 | 40.1 ± 0.2 | 124.6 ± 0.7 |
| Bulk UV exposed Un-annealed W fibre | 8.88 ± 0.05 | 40.7 ± 0.2 | 124.6 ± 0.7 |
| Bulk UV exposed Annealed W fibre | 8.95 ± 0.05 | 40.2 ± 0.2 | 124.7 ± 0.7 |
| Un-annealed W fibre LPG | 8.97 ± 0.05 | 40.4 ± 0.2 | 124.7 ± 0.7 |
| Annealed W fibre LPG | 9.10 ± 0.05 | 39.9 ± 0.2 | 124.6 ± 0.7 |
| Pristine Un-annealed SMF28 | 7.90± 0.05 | / | 124.5 ± 0.7 |
| Pristine Annealed SMF28 | 7.92± 0.05 | / | 124.6 ± 0.7 |
| Bulk UV exposed Un-annealed SMF28 | 7.91± 0.05 | / | 124.4 ± 0.7 |
| Bulk UV exposed Annealed SMF28 | 7.92± 0.05 | / | 124.5 ± 0.7 |
| Un-annealed SMF28 LPG | 7.90± 0.05 | / | 124.6 ± 0.7 |
| Annealed SMF28 LPG | 7.91± 0.05 | / | 124.5 ± 0.7 |

Table 2 shows similar dimensional changes exist in the bulk UV exposed W fibre samples after annealing. There is an increased core diameter and a decreased inner cladding diameter while the pristine W fibre samples and all the SMF28 fibre samples have no obvious change. These results indicate that annealing has a significant effect on the W fibre samples with UV exposure. In the SMF 28 fibre, without an inner cladding no dimensional changes within experimental error are resolved after annealing. Besides the changes due to the annealing, it's worth noting that for the W fibre the initial UV exposed periodic LPG structures have a slightly enlarged core diameter, while the initial bulk UV exposed samples have a larger inner cladding diameter, compared with the pristine fibre. The conclusion is that periodic UV exposure leads to greater core expansion while bulk UV exposure mostly affects the inner cladding. This difference is discussed below.

DISCUSSION

The thermal expansion coefficient, transition temperature and melting temperature of different components in the W fibre are listed in Table 3 [12, 21–25]. The annealing temperature used ($T = 300$ °C) is higher than the transition temperatures of pure boron oxide ($B_2O_3$) and pure phosphorous pentoxide ($P_2O_5$), but lower than that of pure silica ($SiO_2$) and pure germanium dioxide ($GeO_2$). The thermal coefficients of fluorine are not listed, but fluorine doping is known to lead to a significant drop in glass transition [26]. Since the transition temperature generally decreases with increasing dopants [21], and most of the dopants reside in the core and the inner cladding, the transition temperatures of the two layers are likely to be close to or less than the annealing temperature, suggesting they will experience greater changes with annealing. On the other hand, the pure silica outer cladding will not.

Table 3. Thermal expansion coefficient ($α$), transition temperature ($T_g$) and melting temperature ($T_m$) of different components

|  | $SiO_2$ | $B_2O_3$ | $GeO_2$ | $P_2O_5$ | F |
|---|---|---|---|---|---|
| $α$ (×$10^{-7}$/K) | 4.1 | 151 | 64 | 13.8 | / |
| $T_g$ (°C) | ~1173 | ~270 | ~530 | ~264 | / |
| $T_m$ (°C) | 1600 | 450 | 1115 | 340 | / |

The doped components have much larger thermal expansion coefficients than pure silica, and that of $B_2O_3$ and $GeO_2$ are larger than $P_2O_5$, from which it is inferred that the core will have the largest thermal expansion coefficient, while the outer cladding the lowest. These differences in fact give rise to a general formation of tensile stress between the outer cladding and the inner layers for both the bi- and tri-material composite fibres during drawing when originally fabricated. Based on the principle of material superposition, the thermal expansion coefficients of the W fibre core, the inner cladding and the outer cladding are approximated as $α_{core}$ ~ 53.2×$10^{-7}$/K, $α_{inner\ cladding}$ ~ 5.0×$10^{-7}$/K and $α_{outer\ cladding}$ ~ 4.1×$10^{-7}$/K. During the cooling process after fibre drawing from the preform at $T = 1900$ °C, the shrinkage of the core is larger than the inner cladding and the outer cladding has the smallest contraction. The outer cladding consolidates first, followed by the inner cladding and then the core generating tensile stress at the interface between layers.

When exposed to UV light, the tensile stress in the W fibre initially increases as the core glass densifies (Fig. 11). For bulk exposure, the irradiation also leads to inner cladding changes. Since the core does not change in a measurable way, we can infer that the inner cladding is affected more than the core due to a much larger volume (despite lower UV absorption). The inner cladding expands both outwards and inwards with an increasing diameter, compensating the expansion of the core, making the core increase slightly less than it might otherwise. When the periodic structure characteristic of an LPG is fabricated, parts of the fibre are exposed while the other parts are not. The stress increase is only in the exposed regions and not in the unexposed. Further, these unexposed regions likely inhibit expansion of the inner cladding in the exposed regions through a mechanical volume effect allowing the core to expand, explaining the differences in results. This has some resemblance to the way mechanical stress and strain can be used within tri-material systems to control waveguide properties by adjusting layer thicknesses and exploiting differences in expansion coefficient [27]; here, in optical fibre form the inner cladding is the equivalent two-dimensional layer used to adjust mechanical properties and responses.

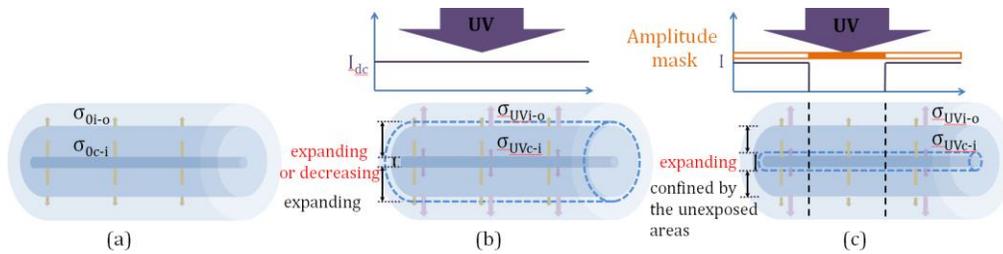

Fig. 11. Schematic diagram of the stress distribution in W fibre samples before annealing: (a) pristine fibre; (b) bulk UV exposed fibre; (c) LPG. ($\sigma_{0c\text{-}i}$/ $\sigma_{0i\text{-}o}$: initial stress at the core-inner cladding interface and inner-outer cladding interface; $\sigma_{UVc\text{-}i}$/ $\sigma_{UVi\text{-}o}$: UV exposure induced stress at the core-inner cladding interface and inner-outer cladding interface; I: UV fluence intensity).

During the post-annealing process, only the core and the inner cladding become softer after the temperature increases over 300 °C, and experience relaxation of the stresses, inherited in the drawn history, which have measurable dimensional changes, while the outer cladding remains comparatively stiff. The average refractive index change of the core and the inner cladding is simulated to be $\Delta n_{core}$ ~ -5×10$^{-4}$ and $\Delta n_{inner\ cladding}$ ~ +5×10$^{-5}$, close to those anticipated from stress relaxation (up to 10$^{-4}$) [28]. The strain relaxation due to annealing effectively reduces the negative material contribution to sensitivity and compensates the positive waveguide contribution. As a result, temperature sensitivity drops dramatically while the strain sensitivity remains near zero.

For SMF 28 fibre, due to the low doping in the core, the transition temperature is not sufficiently low, so thermal annealing has no measurable effect within error of the experiment. As a simple bi-material system, it also lacks the mechanical flexibility a tri-material system such as the W fibre may offer.

The measurable dimension change, an example of real nanoscale glass milling, occurs at a relatively low temperature – this may seem a surprising result in light of many key assumptions made in the grating literature, particularly those based on Bragg grating sensors stated to operate up to 300 °C (though we note telecommunications sets a prescient upper limit of 80 °C for performance over 20 years). In retrospect, we can now summarize three main stages describing this process:

(1) the tri-material W fibre design, including the composition and the waveguide structure, which makes the transition temperature of the core and inner cladding sufficiently low;
(2) the UV exposure, inducing increased tensile stresses at the core/cladding interface which in turn leads to stress relaxation prior to annealing; and
(3) thermal annealing.

All three explain why fibre design and composition are integral. Another key point is the fabrication of the LPG, which is sensitive enough to detect such nanoscale transverse dimensional changes, in comparison to short period fibre Bragg gratings where the process may also be less impactful because of the smaller scale separating exposed and non-exposed regions. The difference between bulk and periodic exposures generally implies that tuning the LPG pitch will tune the processes and could explain the origin of some of the empirical observations in the literature noting a high sensitivity in balancing material and waveguide dispersion. This kind of nanoscale glass milling must also exist in other photonic components such as fibre Bragg gratings and in planar waveguides where stresses are far more pronounced [27,28], although it needs further work to determine the role of much shorter periods. Nonetheless, given that relatively low temperature annealing is commonly used in stabilizing components, this work calls for substantial further study.

CONCLUSIONS

In summary, we have experimentally demonstrated the concept and idea of nanoscale glass milling through fibre design using composite glass systems, irradiation and mild annealing. The LPGs were fabricated in W single mode fibre with almost zero strain insensitivity and a temperature sensitivity of $d\lambda/dT = -0.111$ nm/K, which is suitable for temperature sensing. SEM microscopy assisted by etching was shown to be an invaluable tool capable of imaging and characterizing such nanoscale changes. After annealing at $T \sim 330$ °C for 1 hour, the LPG was both strain and temperature insensitive, accompanied by diameter changes in the core ($\Delta\phi_{core} \sim 0.13 \pm 0.05$ μm) and inner cladding ($\Delta\phi_{inner\ cladding} \sim -0.5 \pm 0.2$ μm) based on SEM measurements. Such environmentally insensitive gratings are ideal for in-fibre filters, bend sensors, mode converters and many other applications, such as ramp filters for attenuation based FBG interrogators. They are particularly useful in amplifier and fibre laser applications where pumping of rare earth dopants, for example, generate heat and elevate core temperatures significantly. The results reported here are the first practical step in using moderate glass milling conditions to fine tune photonic components. The process has also shed more light on the fundamental understanding of basic annealing of glass and glassy composites.